\documentclass[preprint]{revtex4}
\usepackage{graphicx}

\makeatletter

\begin{document}

\title{Synchronization and time shifts of dynamical patterns for mutually delay-coupled fiber ring lasers}

\author{Leah B.~Shaw}
  \affiliation{US Naval Research Laboratory, Code 6792, Nonlinear Systems
  Dynamics Section, Plasma Physics Division, Washington, DC 20375}

\author{Ira B.~Schwartz}
  \affiliation{US Naval Research Laboratory, Code 6792, Nonlinear Systems
  Dynamics Section, Plasma Physics Division, Washington, DC 20375}

\author{Elizabeth A.~Rogers}
  \altaffiliation{Current address:  Electron Physics Group, National Institute of Standards and Technology, Gaithersburg, MD 20899-8412}
  \affiliation{Department of Physics and Institute for Research in Electronics and Applied Physics, University of Maryland, College Park, Maryland 20742}

\author{ Rajarshi Roy}
  \affiliation{Department of Physics and Institute for Research in Electronics and Applied Physics, University of Maryland, College Park, Maryland 20742}
  \affiliation{Institute for Physical Science and Technology, University of Maryland, College Park, Maryland 20742}

\begin{abstract}
A pair of coupled erbium doped fiber ring lasers is used to explore the
dynamics of coupled spatiotemporal systems. The lasers are mutually
coupled with a coupling delay less than the cavity round-trip
time.  We study synchronization between the two lasers in the experiment and in a delay differential equation model of the system. Because the lasers are internally perturbed by spontaneous
emission, we include a noise source in the model to obtain stochastic
realizations of the deterministic equations.   Both amplitude
synchronization and phase synchronization are considered.  We use the
Hilbert transform to define the phase variable and compute phase
synchronization.  We find that synchronization increases with coupling
strength in the experiment and the model.  When the time series from two lasers are time-shifted in either direction by the delay time, approximately equal synchronization is frequently observed, so that a clear leader and follower cannot be identified.  We define an algorithm to determine which laser leads the other when the synchronization is sufficiently different with one direction of time shift, and statistics of switches in leader and follower are studied.  The frequency of switching between leader and follower increases with coupling strength, as might be expected since the lasers mutually influence each other more effectively with stronger coupling.
\end{abstract}

\maketitle

\textbf{The main goal of this paper is to explore the synchronization of mutually
delay-coupled spatiotemporal systems and develop techniques to identify the
evolving phase relationships between them. We illustrate our ideas in the
specific case of mutually coupled fiber ring lasers, which exhibit a remarkable
ability to adapt their dynamics so that much of the time there is no clear
leader or follower, even though signals propagate from one system to the other
with an appreciable time delay.  The lasers are described by a system of
stochastic delay differential equations, so as to include the effect of
spontaneous emission in the erbium doped fiber amplifiers that constitute the
active media.  We use statistical measures to study phase and amplitude synchrony in
model and experimental delay-coupled systems from their time series.}

One fascinating area of dynamics in nature is the study of how systems
respond to each other due to their interactions. When systems behave
dynamically, interactions between them may cause the systems to operate in
a similar, or coherent, manner. If the measured signals are similar enough over
time, the correlated motion is termed synchronized. There now exist
several excellent reviews on synchronization dynamics in the literature,
such as Refs.~\cite{BoccalettiKOVZ02,pikovsky,DingDDGIPSY97,Strogatz00},
and they cover a wide range of applications from many fields of science.

In general, synchronization between interacting systems may be quantified
by examining and comparing the output time series from each dynamical
system.  Although the dynamics may be considered for the general coupling between $N$ systems, in this paper we restrict the number of dynamical systems to $N=2$. Then we may consider only two coupling schemes; either mutual (bidirectional), or unidirectional.

 Suppose $\mathbf{X}_{i}:R\rightarrow R^n,i=1,2$ denote
the vector output time series measured from each dynamical
system. Then several types of synchronization may be classified depending
on the type of coherence measure. The systems are in complete synchronization
if $\mathbf{X}_{1}(t)=\mathbf{X}_{2}(t)$. Complete synchronization occurs in coupled phase oscillators
\cite{Strogatz00} as well as in coupled chaotic oscillators
\cite{FujisakaY83,PecoraC90}. In this case, amplitudes and
phases are identical.

If the amplitudes are uncorrelated but the phases are locked, or entrained,
between the two signals, then the systems are said to be in phase
synchrony \cite{PikovskyRK96}. One other type of synchronization
deals solely with the unidirectional coupling between two oscillators of drive and response type,
and is termed generalized synchronization \cite{RulkovSTA95}. In
generalized synchronization, there exists a functional relationship
between the drive and response, where there
might exist a function $F$ such that $\mathbf{X}_{1}(t)=F(\mathbf{X}_{2}(t)).$
In a more general setting, this may also be thought of as a generalized
entrainment in dynamics, whereby one system is entrained functionally to
another. Many examples of entrained systems occur in singularly perturbed
problems, and specifically in systems with highly different
time scales \cite{SchwartzMBL04}.

If there is a parameter mismatch or noise in the dynamical systems, complete
synchronization may not be possible, and other measures of synchronization
are needed. One possible example of phase synchronization occurs when
the amplitudes are correlated but locked in phase at a value other
than zero. Such a system exhibits lag synchronization when for some
$\tau>0$ we have $\mathbf{X}_{1}(t)=\mathbf{X}_{2}(t-\tau).$ That
is, the two outputs of the coupled dynamical systems appear shifted
in time with respect to one another. For mutually coupled chaotic
systems such as  Rossler attractors that are mismatched in frequency, lag synchronization is one of the routes to complete synchrony as coupling
is increased \cite{RosenblumPK97}.  It should be
noted that lag synchronization may occur without the presence of delay
in the coupling terms.

On the other hand, if delay is introduced into the coupling terms to model finite time
signal propagation, then synchronized behavior may still occur.
When there is a clear time lag between the delay-coupled
dynamics, the systems are said to exhibit achronal synchronization.
Achronal synchronization exhibits a clear leading time series which
is followed by a lagging time series. Heil \textit{et al.}~\cite{HeilFEMM01}
showed the existence of achronal synchronization in a delay-coupled
semiconductor laser experiment, as well as in a single mode model of the
delay-coupled lasers in which stochastic effects modeling spontaneous
emission are included. The time series shows a clear leader 
with delay equal to the coupling delay time. Other groups have also
considered leader-follower synchronization
in single-mode semiconductor equations \cite{WhiteMM02}.

One of the non-intuitive facets of interacting systems which synchronize
with delay is that of anticipation in systems with short time delay.
First observed in unidirectionally coupled systems, in contrast to lag synchronization,
anticipatory synchronization occurs when a response in a system's
state is replicated not simultaneously but anticipated by the response
system \cite{Voss00,Voss01,TangLiu}. An example of anticipation in synchronization
is found in coupled semiconductor lasers \cite{Masoller01}. Here,
the author followed Ref.~\cite{Voss00} in the design of a unidirectional coupling
arrangement, in which two single mode semiconductor lasers with delayed
optical feedback and delayed injection coupling were modeled. 
Cross-correlation
statistics between the two intensities showed clear maxima at delay
times consisting of the difference between the feedback and the coupling
delay. Anticipatory responses in the presence of stochastic drives,
equally applied to transmitter and receiver, have been observed in models of
excitable media as well \cite{CiszakCMMT03}.

Given that both lag and anticipatory dynamics may be observed in delay-coupled systems which are deterministic, it is natural to ask whether
the systems may exhibit coexisting features.  In single
mode mutually coupled lasers, this is indeed the case \cite{WuZ03}.
In fact, in the absence of noise (the authors include a noise term
in their model, but turn it off for the simulations), switching between
leading and following state is observed.
In a similar model with spontaneous
emission included as a noise source, theory and experiment have exhibited achronal synchronization
\cite{MuletMHF04}, with switching between leader and follower. The authors conjecture
that for semiconductor lasers, changes in leader-follower roles may
occur during sharp dropouts of the laser intensity.

In the above discussion, it is clear that the role of signal propagation
time is important in the dynamics of leader and follower in coupled
systems. When noise is sufficiently
large, it is not easy to detect changes in leader and follower
statistically. This is especially the case when the systems are
  stochastic and close to a synchronized state.
In Ref.~\cite{MuletMHF04}, the authors compute statistics on the phase differences
and show that they might correspond to a random walk. In this paper,
we develop methods to extract statistical behavior of the switches
in leader and follower in mutually coupled fiber lasers. 

The chaotic dynamics of fiber ring lasers have been studied in the
past.
 An experiment on the coupling between polarization modes
was set up and modeled using delay differential equations in
\cite{WilliamsR96}. Other experiments on synchronization with fiber lasers have been
reported in \cite{VanWRoy,WangS02,ImaiMI03}, and noise-induced generalized synchronization
in fiber ring lasers has been reported in \cite{DeShazerTKR04}.
 Modeling the ring laser yields a system
of equations which consists of coupled difference and differential
delay equations. To obtain better agreement
with experiment, it was found that inclusion of spontaneous emission effects was
necessary in the modeling, which resulted in  a stochastic
difference-differential system of equations \cite{WilliamsGR97}, and it is this approach we follow here.

For mutual coupling of two ring lasers with delay,  we analyze both experimental results and a delay differential equation model of the system.  Synchronization with delay occurs and increases with coupling strength.  Generally, approximately equal synchronization is observed between the two lasers when time-shifted in either direction by the delay time, so that a clear leader and follower cannot be identified.  We define an algorithm to determine which laser leads the other when the synchronization is sufficiently different, and statistics of switches in leader and follower are studied.

We introduce the experimental setup and corresponding model in Sec.~\ref{sec:experiment}-\ref{sec:model}.  The dynamics of the system are outlined in Sec.~\ref{sec:dynamics}.  Synchronization of the coupled lasers is discussed in Sec.~\ref{sec:synchronization} and switching of leader and follower in Sec.~\ref{sec:switching}.

\section{Experimental methods}
\label{sec:experiment}

The experimental setup consists of two erbium doped fiber ring lasers
(EDFRLs) coupled with two passive coupling lines.  Each EDFRL has
approximately \mbox{17 m} of erbium-doped fiber, the active medium,
and approximately \mbox{29 m} of passive single (transverse) mode fiber,  making the total length of each cavity approximately \mbox{46 m}.  The lengths of the cavities are matched within \mbox{1 cm} of each other.  Though the doping density and the lengths of the active media are the same in both lasers, defects and imperfections in the fiber make the lasers nonidentical.  The erbium ions in the active medium are pumped with identical \mbox{980 nm} semiconductor lasers at a pump power of \mbox{120 mW}.  The lasing threshold for both lasers is approximately \mbox{20 mW}. This results in approximately \mbox{1 mW} of light circulating within each ring.  An optical isolator is inserted within each ring cavity to ensure unidirectional propagation within the rings.   

\begin{figure}
\includegraphics[
 width=2.75in,
 keepaspectratio]{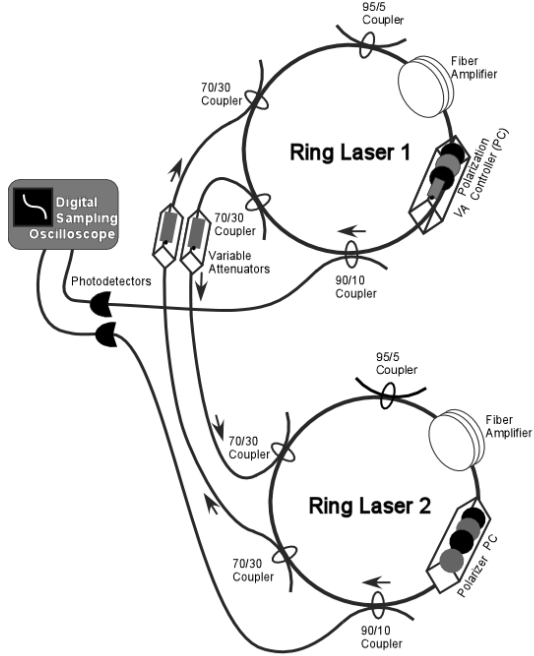} 
\caption{Experimental setup for the coupled fiber laser system.}
\label{fig:expsetup}
\end{figure}

Each laser contains 4 fiber-optic evanescent field couplers placed at similar locations in both rings.  These consist of two 70/30 couplers which input and output light between the lasers, one 90/10 coupler for monitoring, and a 95/5 coupler as an extra port for applications that are beyond the scope of this paper.  The locations of the couplers are shown in Figure \ref{fig:expsetup}.  The ports of the couplers not in use are angle cleaved to ensure that there are no back reflections and were monitored to ensure that light was propagating in the correct direction within the cavities.

The lasers are connected via two injection lines, which consist of passive single mode optical fiber, one splitter, and a variable attenuator.  In this configuration, we have the ability to monitor and control the injection amplitude between the lasers, through the splitter and variable attenuator respectively, and to observe it on an oscilloscope.  The coupling strength is defined to be the ratio of the power of light in the injection line to the power of light in the source ring.  (Note that both lasers have been adjusted to have the same power.)  The lowest coupling strength that can be resolved in our system is 0.0001.  The injection lines are approximately \mbox{9 m} long, corresponding to a travel time between the two lasers of approximately \mbox{45 ns}, and again they are matched within \mbox{1 cm}.  In the experiments presented here, the two coupling strengths are always the same (symmetric coupling), though the electric field from each laser undergoes different phase and polarization changes due to fiber imperfections along their separate paths.  

The electric field intensity of each laser is monitored using a \mbox{125 MHz} bandwidth photodetector and a \mbox{1 GSample/s} digital oscilloscope.  The optical spectrum of each laser is also monitored, and the spectra of the uncoupled lasers are matched to within \mbox{1 nm}.

\section{Description of model}
\label{sec:model}

We use the model of \cite{WilliamsGR97} with the modifications introduced in \cite{Rogersthesis, Rogersinprep} for coupled fiber ring lasers.  Only a single polarization mode in each laser is considered here.  The model is characterized by the total population inversion $W(t)$ (averaged over the length of the fiber amplifier) and the electric field $E(t)$ in each laser.  Two delay times occur in the model:  $\tau_R$, the cavity round-trip time, and $\tau_d$, the delay in the coupling between lasers.  The equations for the model dynamics are as follows:
\begin{eqnarray}
E_j(t) &=& R \exp \left[\Gamma (1-i\alpha_j) W_j(t)+i\Delta \phi\right] E^{\text{fdb}}_j(t) \nonumber \\
&& +\xi_j(t) \label{Eequ} \\
\frac{dW_j}{dt} &=& q -1-W_j(t) \nonumber \\
&& -\left|E^{\text{fdb}}_j(t)\right|^2 \left\{ \exp\left[2\Gamma W_j(t)\right]-1 \right\}, \label{Wequ}
\end{eqnarray}
where
\begin{equation}
E^{\text{fdb}}_{1,2}(t)=E_{1,2}(t-\tau_R)+\kappa E_{2,1}(t-\tau_d). \label{Efdb}
\end{equation}
$E_j(t)$ is the complex envelope of the electric field in laser $j$, measured at a given reference point inside the cavity.  $E^{\text{fbd}}_j(t)$ is a feedback term that includes optical feedback within laser $j$ and optical coupling with the other laser.  Time is dimensionless, measured in units of the decay time of the atomic transition, $\gamma_{||}$.  The active medium is characterized by the dimensionless detuning $\alpha_j$ between the transition and lasing frequencies and by the dimensionless gain $\Gamma=\frac{1}{2}a L_a N_0$, where $a$ is the material gain, $L_a$ the active fiber length, and $N_0$ the population inversion at transparency.  The ring cavity is characterized by its return coefficient $R$, which represents the fraction of light remaining in the cavity after one round-trip, and the average phase change $\Delta \phi=2\pi n L_p/\lambda$ due to propagation of light with wavelength $\lambda$ along the passive fiber of length $L_p$ and index of refraction $n$.  Energy input is given by the pump parameter $q$.  The electric field is perturbed by a complex Gaussian noise source $\xi_j$ with standard deviation $D$.  Coupling between the lasers is characterized by the coupling strength $\kappa$, which was varied from 0 to 0.014.  The same $\kappa$ is used in both lasers (symmetric mutual coupling).  Values of the parameters in the model, similar to those in the experimental system, are given in Table \ref{tab:parameters}.

\begin{table}
\caption{\label{tab:parameters}Parameters used in the coupled fiber laser model.}
\begin{ruledtabular}
\begin{tabular}{cccl}
Parameter & Value & Units & Description \\
\hline

$R$ & 0.4 & & output coupler return coefficient \\
$a$ & $2.03 \times 10^{-23}$ & m$^2$ & material gain coefficient \\
$L_a$ & 15 & m & length of active fiber \\
$L_p$ & 27 & m & length of passive fiber \\
$N_0$ & $10^{20}$ & m$^{-3}$ & transparency inversion \\
$\Gamma$ & 0.0152 & & dimensionless gain \\
$\alpha_1$ & 0.0352 & & detuning factor, laser 1 \\
$\alpha_2$ & 0.0202 & & detuning factor, laser 2 \\
$n$ & 1.44 & & index of refraction \\
$\lambda$ & $1.55\times 10^{-6}$ & m & wavelength \\
$\Delta \phi$ & $1.58\times 10^8$ & & average phase change \\
$D$ & 0.02 & & standard deviation of noise \\
$q$ & 100 & & pump parameter \\
$\gamma_{||}$ & 100 & s$^{-1}$ & population decay rate \\
$\tau_R$ & $201.6\times 10^{-9}$ & s & cavity round-trip time \\
$\tau_d$ & $45\times 10^{-9}$ & s & delay time between lasers \\
$\kappa$ & 0-0.014 & & coupling strength
\end{tabular}
\end{ruledtabular}
\end{table}

Eqns.~\ref{Eequ}-\ref{Wequ} consist of a delay differential equation for $W(t)$ coupled to a map for $E(t)$.  We integrated Eqn.~\ref{Wequ} numerically using Heun's method.  The time step for integration was $\tau_R/N$, where $N=600$.  This step size corresponds to dividing the ring cavity into $N$ spatial elements.

Because of the feedback term $E^{\text{fdb}}_j(t)$ in Eqn.~\ref{Eequ}, one can think of Eqn.~\ref{Eequ} as mapping the electric field on the time interval $[t-\tau_R,t]$ to the time interval $[t,t+\tau_R]$ in the absence of coupling ($\kappa=0$).  Equivalently, because the light is traveling around the cavity, Eqn.~\ref{Eequ} maps the electric field at all points in the ring at time $t$ to the electric field at all points in the ring at time $t+\tau_R$.  We can thus construct spatiotemporal plots for $E(t)$ or the intensity $I(t)=\left| E(t) \right|^2$ by unwrapping $E(t)$ into segments of length $\tau_R$.

The experimental measurements were of the intensity of light from each laser after passing through a 125 MHz bandwidth photodetector.  To correspond with the experiment, we computed intensities from model and applied a low pass filter with $f_0=125$ MHz, multiplying the Fourier transform by the transfer function
\begin{equation}
G=\left\{ \left(i \frac{f}{f_0}+1\right) \left[-\left(\frac{f}{f_0}\right)^2+i \frac{f}{f_0}+1\right] \right\} ^{-1}.
\end{equation}
It should be noted that the time step for integration is an order of magnitude smaller than the smallest timescale allowed by the filtering function.  We expect that the results are insensitive to the integration time step as long as it is sufficiently small.  Test runs with a finer mesh, $N=1200$ time steps per round-trip, yielded similar behavior to that described here.

All subsequent analysis of the model properties is based on the filtered intensity.

\section{Effect of coupling on system dynamics}
\label{sec:dynamics}

In this section we discuss the types of dynamics observed in simulations and the experimental system.  After transients have died out, each laser settles into a pattern that is approximately repeated every round-trip.  The pattern may shift or change over time intervals of tens or hundreds of round-trips.  Due to the presence of noise in the model, even the uncoupled lasers ($\kappa=0$) are not fully periodic for these parameter values.

\begin{figure*}
\includegraphics[
 width=6.5in,
 keepaspectratio]{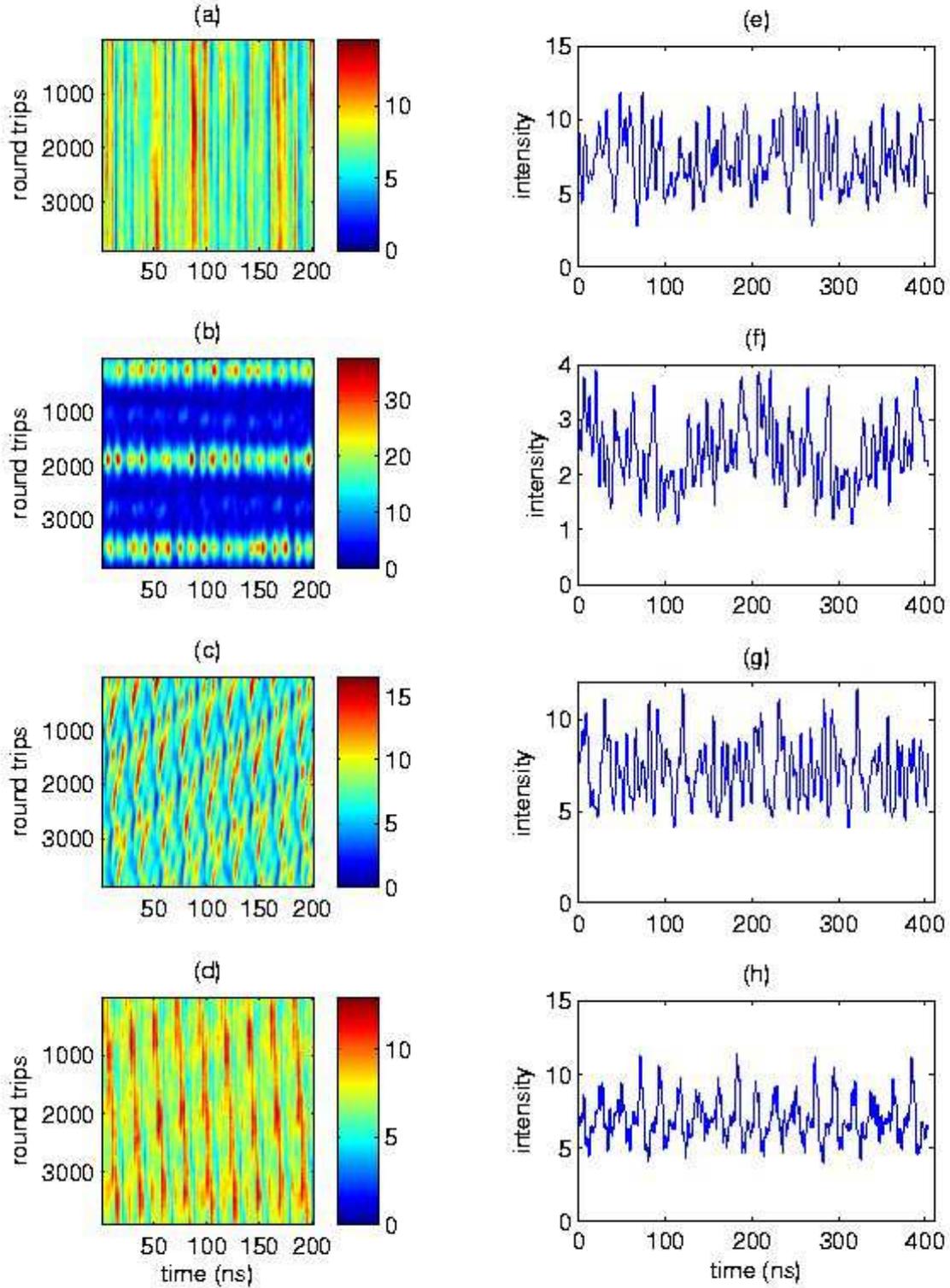} 
\caption{(a)-(d):  Spatiotemporal plots of Laser 1 intensity for the model.  [(a) $\kappa=0$, (b) $\kappa=0.005$, (c) $\kappa=0.009$, (d) $\kappa=0.011$.]  (e)-(h):  Model intensity vs.~time for two round-trips.  [Same $\kappa$ values as in (a)-(d).  Note that (f) shows typical behavior in the low intensity regions of the pulsing regime.]}
\label{fig:ST_sim}
\end{figure*}

Typical spatiotemporal plots and time traces for the model are shown in Figure \ref{fig:ST_sim}.  Spatiotemporal plots are constructed by displaying each round-trip as a row in the diagram, colored according to the laser intensity, with subsequent round-trips forming subsequent rows.

The system displays approximately steady behavior for small coupling ($\kappa \leq 0.003$) (Fig.~\ref{fig:ST_sim}a).  As the coupling increases ($\kappa>0.003$), it then enters a regime where pulsing occurs, in which the intensity alternates between pulsing and dropping to a low value while the inversion builds up (Fig.~\ref{fig:ST_sim}b).  Although the spatiotemporal plot may appear more periodic during the pulses, the intensity is actually highly irregular.  When the coupling increases further ($\kappa \geq 0.007$), the system leaves the pulsing regime and displays complicated behavior.  Traveling wave solutions are commonly observed (e.g., Fig.~\ref{fig:ST_sim}c).  As the coupling increases, the traveling waves become less prominent (e.g., Fig.~\ref{fig:ST_sim}d).  Time traces of the intensity are also shown (Figure \ref{fig:ST_sim}e-h) for two adjacent round-trips.  Approximate repetition of the intensity pattern from one round-trip to the next can be seen.

\begin{figure*}
\includegraphics[
 width=7in,
 keepaspectratio]{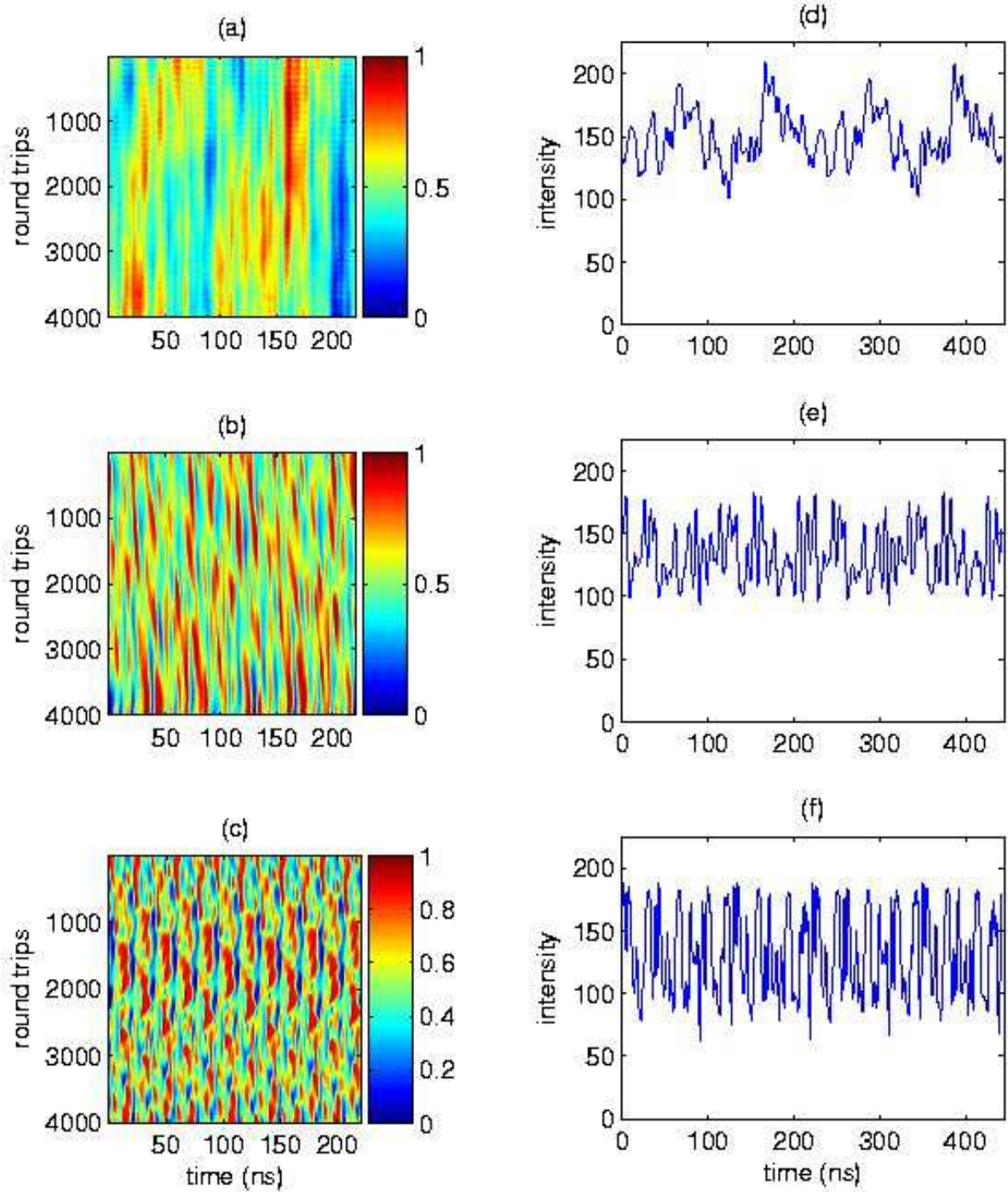} 
\caption{(a)-(c):  Spatiotemporal plots of Laser 1 intensity for experiment (intensities rescaled to have equal ranges).  [(a) $\kappa=0$, (b) $\kappa=0.000285$, (c) $\kappa=0.0228$.]  (d)-(f):  Experimental intensity vs.~time for two round-trips.  [Same $\kappa$ values as in (a)-(c).]}
\label{fig:ST_exp}
\end{figure*}

Spatiotemporal plots for the experiment are shown in Figure \ref{fig:ST_exp}.  Because the round-trip time is not an integer multiple of the 1 ns sampling time, spacetime plots were constructed by the following procedure.  The experimental intensity time trace was expanded to 10 times as many time points by linear interpolation.  Subsequent round-trips were aligned by shifting the relative position of the rows to maximize correlation between the rows.  Spatiotemporal plots for the experiment are normalized so that the intensities range from 0 to 1.

The experimental system displays similar behavior to the model.  When $\kappa=0$, approximately steady behavior is observed (Figure \ref{fig:ST_exp}a).  When very small coupling is turned on, higher frequency structures emerge in the spatiotemporal pattern (Figure \ref{fig:ST_exp}b).  Complicated spatiotemporal patterns are seen at stronger coupling (Figure \ref{fig:ST_exp}c).  Pulsing was not observed in this experimental study.  However, pulsing in other fiber ring laser systems has been commonly reported in the literature (e.g., \cite{BielawskiD95}).  As in the model, the experimental time traces show the approximate repetition of the intensity pattern in subsequent round-trips (Figure \ref{fig:ST_exp}d-f).

\section{Synchronization}
\label{sec:synchronization}

\begin{figure}
\includegraphics[
 width=3.25in,
 keepaspectratio]{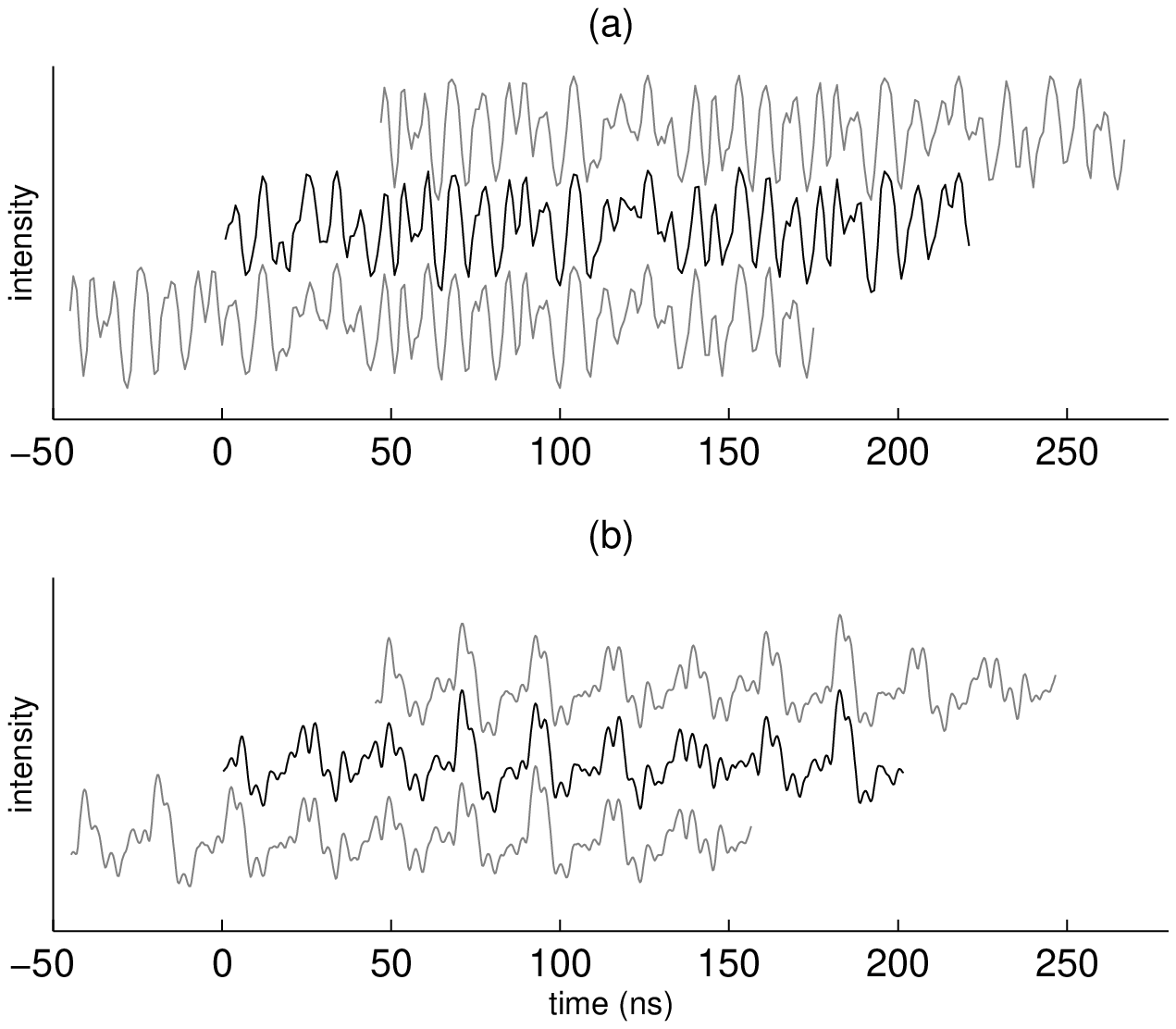} 
\caption{Time traces of intensity (arbitrary units).  In each plot, the center bold curve is intensity vs.~time for a single round-trip in Laser 1.  The lighter curves above and below it are intensity vs.~time for the same time interval in Laser 2, but plotted with the curve shifted forwards or backwards in time by the delay time $\tau_d$.  The correspondence between the bold and light curves indicates synchrony between the two lasers with a shift of $\tau_d$.  (a) Experimental results for $\kappa=0.0114$, (b) model results for $\kappa=0.011$.}
\label{fig:synctimetraces}
\end{figure}

We next consider synchronization between the two lasers.  Because the lasers are mutually delay-coupled, with a delay time $\tau_d$, we expect the coupling to cause lag synchronization in the lasers so that $I_1(t)=I_2(t-\tau_d)$ or $I_2(t)=I_1(t-\tau_d)$.  In Figure \ref{fig:synctimetraces}, we display time traces of the two lasers with an offset of $\tau_d$.  For both the experiment and the model, synchronization can be seen with a time shift in \textit{either} direction.

We studied the synchronization quantitatively using two metrics, the \textit{cross-correlation} $C$ for amplitude synchronization and the \textit{mean phase coherence} $R$ for phase synchronization.  Let $C_{12}$ denote the correlation computed for Laser 1 leading Laser 2 and $C_{21}$ the correlation for Laser 2 leading Laser 1.  Thus for time series of length $K$ time points,
\begin{equation}
C_{ab} = \frac{1}{\sigma_a \sigma_b} \frac{1}{K} \sum_{j=1}^K { \left[I_a (t_j-\tau_d) -\left<I_a\right>\right] \left[ I_b(t_j ) -\left< I_b \right> \right] },
\end{equation}
where $\left< \cdot \right>$ denotes averaging and $\sigma_a,\sigma_b$ are standard deviations in intensity for Lasers $a$ and $b$.

We define $R_{12}$ and $R_{21}$ similarly.  The phase of a time series is defined as the Hilbert phase $\phi(t)$, computed from the Hilbert transform of the intensity.  The mean phase coherence for two time series of length $K$ is then defined as follows \cite{kuramoto,CarrS98,Sowa}:
\begin{equation}
R_{ab}=\left| \frac{1}{K} \sum_{j=1}^K {e^{\Delta \phi_{ab,j}}}  \right|,
\end{equation}
where
\begin{equation}
\Delta \phi_{ab,j}=\phi_a(t_j -\tau_d)-\phi_b(t_j)
\end{equation}
is the phase difference between the $j$th points of the time series when Laser $a$ leads Laser $b$.  $R$ ranges between 0 (no synchronization) and 1 (complete phase synchronization).

\begin{figure}
\includegraphics[
 width=3.25in,
 keepaspectratio]{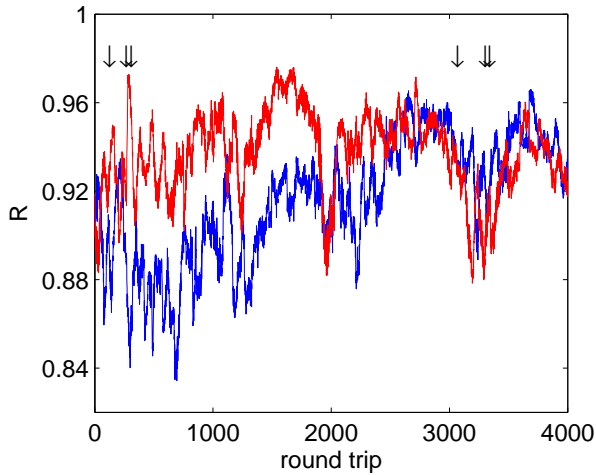} 
\caption{Phase synchrony vs.~round-trip number for experiment with $\kappa=0.0114$ coupling.  Blue curve:  synchrony computed with Laser 2 leading ($R_{21}$); red curve:  synchrony computed with Laser 1 leading ($R_{12}$).  Locations of switches identified by our algorithm are indicated by arrows.}
\label{fig:sync_vs_t}
\end{figure}

The synchronization metrics were calculated for each round-trip, with each laser leading.  We divided the time series into round-trips and shifted each short time series of length $\tau_R$ by the delay $\tau_d$ in either direction.  (Hanging ends were omitted, so that only $\tau_R-\tau_d$ was used in the calculation.)  $C$ and $R$ were then computed.  When analyzing the experimental data, we compensated for the discrete sampling by using linear interpolation to expand the data to 10 times as many time points and then calculating $C$ and $R$ for a 2.5 ns range of coupling delays (centered around the estimated $\tau_d$).  The maximal $C_{12}$, $C_{21}$, $R_{12}$, and $R_{21}$ values for each round-trip (maximized over possible coupling delays) were recorded.  Sample results for phase synchrony are shown in Figure \ref{fig:sync_vs_t}.  The synchronization fluctuates over time, and there are changes in which laser leading gives the maximum synchronization.  The amplitude synchronization behaves similarly.

\begin{figure*}
\includegraphics[
 width=7in,
 keepaspectratio]{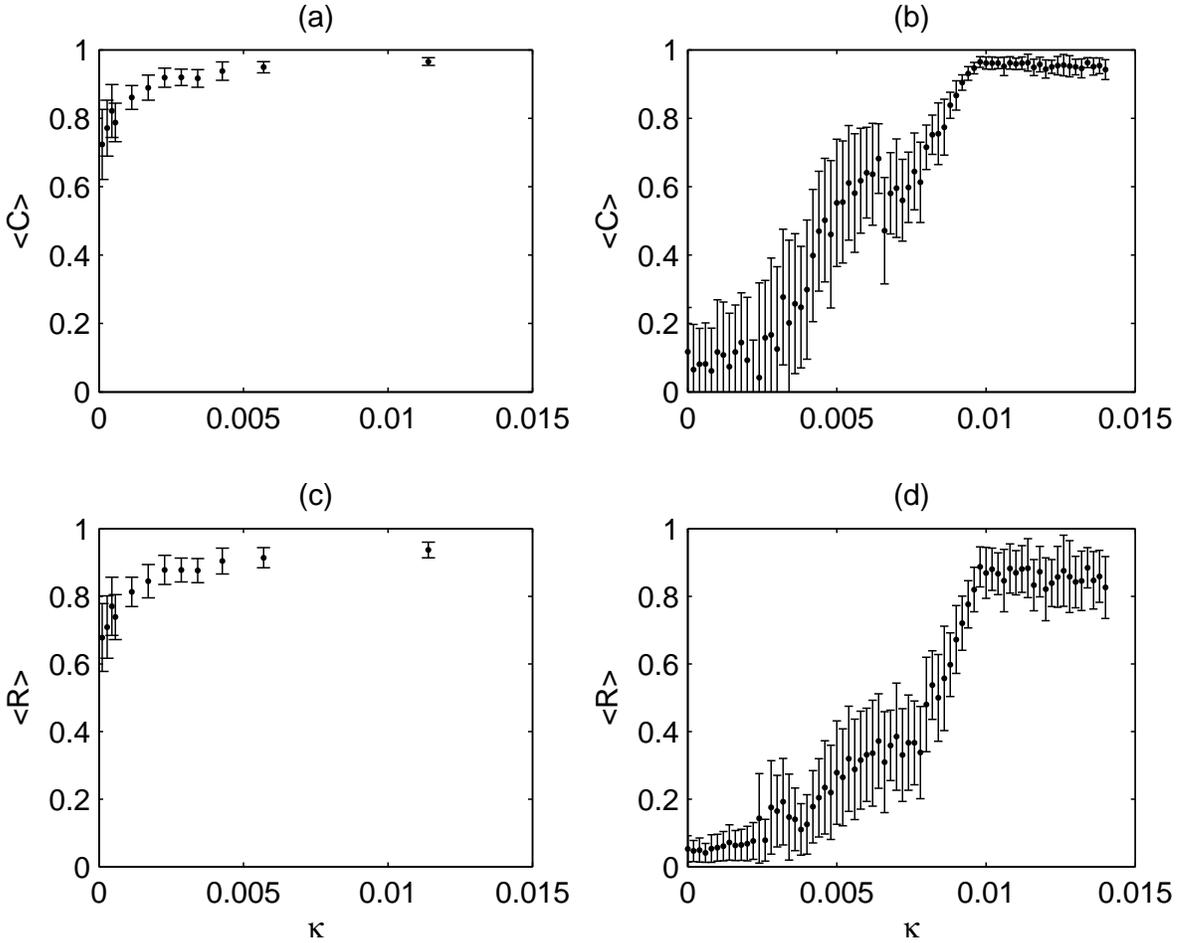} 
\caption{Average correlation between lasers vs.~coupling $\kappa$ for experiment (a) and model (b) and average phase synchrony between lasers vs.~$\kappa$ for experiment (c) and model (d).  Error bars indicate standard deviation over round-trips.}
\label{fig:sync_vs_kappa}
\end{figure*}

To determine the mean amplitude synchronization $\left<C\right>$ for a given coupling value, we take the maximum of $C_{12}$ and $C_{21}$ for each round-trip and then average over all round-trips.  We similarly define the mean phase synchronization $\left<R\right>$ over all round-trips.  Mean synchronizations were computed for each value of the coupling constant $\kappa$.  Experimental data were averaged over 5-10 separate runs of 4500 round-trips each.  Simulations were averaged over 6 runs of $4\times10^4$ round-trips each (with runs separated by at least 0.1 s of simulated time to insure independence).  Results are shown in Figure \ref{fig:sync_vs_kappa}.  Error bars are the standard deviations in $C$ and $R$ over all round-trips, so they are a measure of fluctuations in synchrony.

In both the model and experiments, amplitude and phase synchronization increase as the coupling increases.  However, the shapes of the synchronization vs.~coupling curves are different.  The experimental system is fairly well synchronized even at the smallest nonzero coupling, while the model synchronizes more gradually.  The model also exhibits more fluctuations in synchrony, as exemplified by the larger standard deviations.  For both model and experiment, the amplitude synchronization (cross-correlation) is typically larger than the phase synchronization.

\section{Switching}
\label{sec:switching}

\begin{figure}
\includegraphics[
 width=3.25in,
 keepaspectratio]{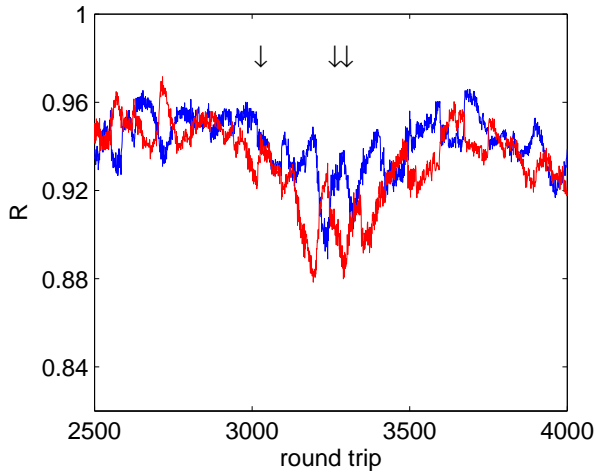} 
\caption{Phase synchrony vs.~round-trip number for experiment with 0.0114 coupling.  Blue curve:  synchrony computed with Laser 2 leading ($R_{21}$); red curve:  synchrony computed with Laser 1 leading ($R_{12}$).  Locations of switches identified by our algorithm are indicated by arrows.}
\label{fig:sync_vs_t_zoom}
\end{figure}

We next consider which laser leads the other during synchronization, and switches in the leader/follower.  In Figure \ref{fig:sync_vs_t}, the two curves are the synchrony computed with Laser 1 leading ($R_{12}$) and with Laser 2 leading ($R_{21}$).  Examining the figure qualitatively, we see that at certain times the synchrony is clearly greater with one laser leading than with the other (e.g., round-trips 1300-1900), and at other times it is more difficult to determine which laser is leading (e.g., round-trips 2500-2900) because the difference in synchrony with time shifts in either direction is small compared to the fluctuations in the synchrony.  The latter portions of Figure \ref{fig:sync_vs_t} are enlarged in Figure \ref{fig:sync_vs_t_zoom}.  We develop an algorithm to compute the leader and follower.

\begin{figure}
\includegraphics[
 width=3.25in,
 keepaspectratio]{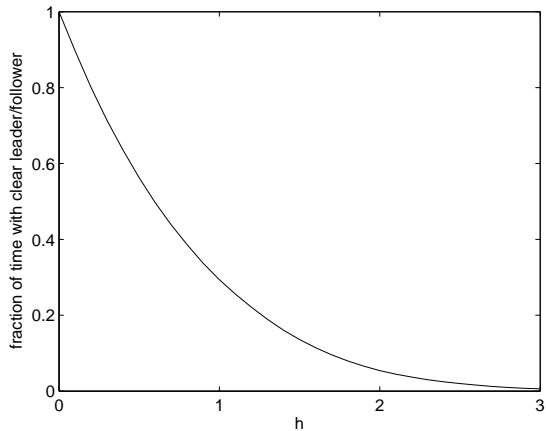} 
\caption{Fraction of round-trips for which leader and follower can be identified vs.~number of standard deviations used in cutoff ($h$) for model with $\kappa=0.01$.}
\label{fig:cutoff}
\end{figure}

For each round-trip, we compute the synchronization with either laser leading.  Let $\Delta C_i = C_{12}(i)-C_{21}(i)$ be the difference between the correlations for each round-trip $i$ within a run, and let $\sigma_C$ be the standard deviation of the $\Delta C_i$.  When $\left| \Delta C_i \right| > h \sigma_C$ for some cutoff factor $h$, we say that the synchronization is substantially greater with one laser leading than with the other, and a clear leader and follower can be identified.  Leader and follower for phase synchronization are defined similarly.  Figure \ref{fig:cutoff} shows the fraction of round-trips for which a leader and follower can be identified for a range of cutoff factors $h$.  Similar curves are obtained for the model and experiment for most values of $\kappa$.  We select $h=1$ to use as the cutoff factor in this study.  For most values of $\kappa$, for both the model and experimental data, the choice $h=1$ leads to approximately 30\% of round-trips having an identified leader and follower for amplitude and phase synchronization.

After leader and follower have been computed, we next search for switches.  The following algorithm is used to identify changes in leader and follower. 
Let $\left\{n_i \right\}$ be the set of round-trip numbers $n_i$ for which  $\left| \Delta C_{n_i} \right| > h \sigma_C$.  Whenever the sign of  $\Delta C_{n_i}$ is opposite the sign of $\Delta C_{n_{i-1}}$, a switch is identified at round-trip $n_i$.

To obtain statistics over several runs, let $N_t$ be the total number of round-trips and $N_{sw,C}$ total number of switches in amplitude synchronization observed.  Then the average number of round-trips between switches is $T_C=N_t/N_{sw,C}$.  We assume square root error in the number of observed switches, leading to an error of $N_t/N_{sw,C}^{3/2}$ in the number of round-trips between switches.  Switches in phase synchronization are located by the same procedure.  An example of switching is given in Figure \ref{fig:sync_vs_t} and Figure \ref{fig:sync_vs_t_zoom}:  locations of switches identified by our algorithm are marked by arrows.  Note that switches are defined only when the leader and follower change; for example, no switch is identified around round-trip 2000 when $R_{12}$ and $R_{21}$ come together with no defined leader and then separate with the same leader and follower as before.

\begin{figure*}
\includegraphics[
 width=7in,
 keepaspectratio]{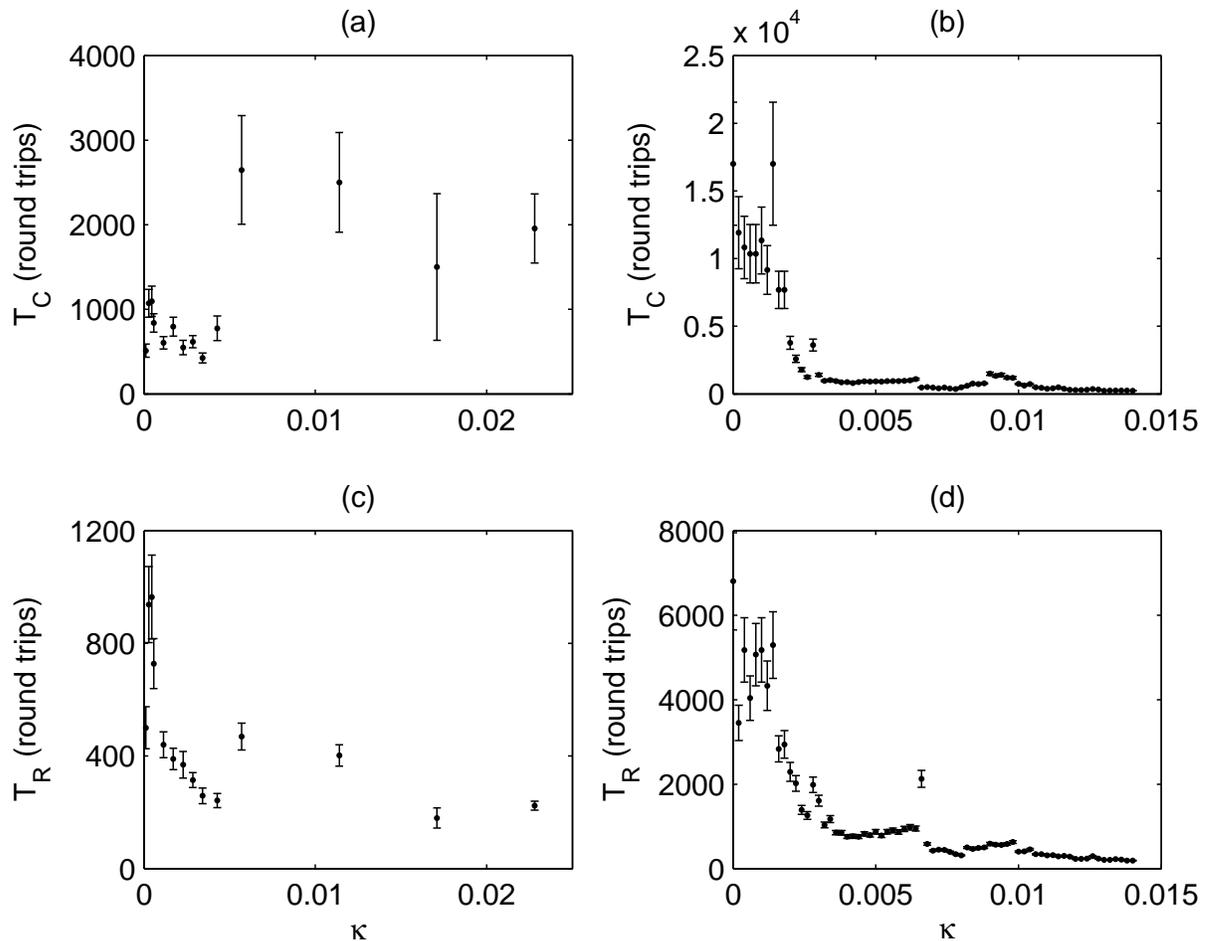} 
\caption{Average number of round-trips between switches in leader/follower as determined by cross-correlation [(a) experiment, (b) model] and by phase synchrony [(c) experiment, (d) model].}
\label{fig:switches}
\end{figure*}

Switches were studied in the model and experiment for a range of coupling strengths, using the same data sets as in Section \ref{sec:synchronization}.  Results are shown in Figure \ref{fig:switches}.  We find that the time between switches in phase synchronization decreases as the coupling increases for both the experiment and model.  Results for the amplitude synchronization are less clear.  For the model, switching in amplitude synchronization follows the same trend as for phase synchronization.  However, the experimental system exhibits fewer switches in amplitude synchronization at larger coupling strengths.

\section{Discussion and conclusions}

We compared experimental measurements with model predictions for a pair
of mutually coupled fiber ring lasers as the coupling strength was
varied. The coupling has similar effect on the dynamics in both model
and experiment. Quantitative measures of amplitude and phase
synchronization  were made, the latter by using the Hilbert
phase.  Approximately equal synchronization was found for forward and reverse time shifts equal to the coupling delay.
 We notice that the lasers
spend considerable periods of time in synchronized states with no
discernable leader or follower, since the synchronization values are very similar for time shifts in either direction.  This is in contrast to the mutually coupled system with delay but no optical feedback described
in \cite{MuletMHF04}, where leader and follower could be identified when the cross-correlation was computed for a short time series.  We defined an algorithm to locate subtle switches
of leader/follower between the two lasers. Our statistical measures allow a leader and
follower to be identified when the synchronization is sufficiently larger
in one direction of the time shift. 

Although the trend of increasing synchronization with increasing coupling was consistent between model and experiment, the shape of the synchronization vs.~coupling curves differed, with the experimental system synchronizing more quickly.  In addition, the fluctuations in synchronization were generally larger in the model.  The model we have used includes only one polarization mode in each laser.  Future extensions to the model should include both polarization modes. This may explain the discrepancy
between the experimental observations and computational prediction in
the shapes of Figure \ref{fig:sync_vs_kappa}.

In \cite{MuletMHF04}, the authors note that for long time series, forward and reverse time shifts yielded similar cross-correlations, although leader and follower could be identified from cross-correlations of short time series.  Our cross-correlations were computed over the natural time scale of a round-trip, and similar results are obtained for shorter time series with length equal to the delay time (data not shown).  Our observation that the system often does not have a clear leader and follower seems robust for reasonable choices of window size for averaging, although further study of appropriate window size may be conducted.

The method for choosing the cutoff to identify clear regions of leader
and follower is based on an internal definition for a given time
series. The choice of the cutoff factor $h$ to be one standard deviation is
somewhat arbitrary, but preliminary results indicate that the trends of the switching
frequency vs.~coupling are preserved for different cutoff factors.  The algorithm for identifying switches may in the future be applied to other systems in which the difference in synchronization for forward and reverse time shifts is more substantial.

We typically observe a higher frequency of switching at higher coupling
values. There are very similar trends in the frequency of switches of
leader/follower for the phase synchronization in experiment and
theory.  This effect might be expected since the lasers mutually influence each other more effectively with stronger coupling and thus can respond to each other more quickly.  However, we cannot explain the differences in corresponding plots
for the amplitude synchronization.  Further study of this phenomenon is needed.

In this study, we considered variations in the coupling strengths only.  The noise strength for the numerical computations was chosen somewhat
arbitrarily. The main guiding principle was to have a reasonable
correspondence between the uncoupled lasers in theory and experiment.  Future studies will consider changes in the noise strength.  Another parameter to be varied is the detuning between lasers, because its value is not known exactly for the experimental system.  The inclusion of a second polarization mode will introduce another set of parameters whose variation may also be studied.

We are indebted to Jordi Garcia Ojalvo for developing the original version of the model and the simulation code, and we acknowledge David DeShazer's assistance in generating spatiotemporal plots of experimental data.  This research was supported by the Office of Naval Research.  LBS is currently a National Research Council post doctoral fellow.  EAR was supported in part by a National Science Foundation fellowship.

\end{document}